\documentclass[10pt]{article}
\usepackage{multicol}
\usepackage{amsmath}
\usepackage{cite}
\usepackage{bm}
\usepackage{amsfonts}
\usepackage{indentfirst}
\newcommand{\ct}[1]{{\textsuperscript{{\cite{#1}}}}}

\newcommand{\bee}{\begin{equation}}
\newcommand{\ee}{\end{equation}}
\newcommand{\beea}{\begin{eqnarray}}
\newcommand{\eea}{\end{eqnarray}}
\newcommand{\vp}{{\bm p}}
\newcommand{\vq}{{\bm q}}
\newcommand{\vx}{{\bm x}}
\newcommand{\vy}{{\bm y}}

\newcommand{\sd}[1]{#1 \hspace{-0.4em} /}

\begin{document}
\title{Kramers-Kr\(\ddot{\mbox{o}}\)nig relation of graphene conductivity}
\author{ {Daqing Liu and Shengli Zhang} \\
        {\small Department of Applied Physics, Xi'an Jiaotong university, Xi'an, 710049, China}\\
       }
\date{}
\maketitle
\begin{abstract}

Utilizing a complete Lorentz-covariant and local-gauge-invariant
formulation, we discuss graphene response to arbitrary external
electric field. The relation, which is called as
Kramers-Kr\(\ddot{\mbox{o}}\)nig relation in the paper, between
imaginary part and real part of ac conductivity is given. We point
out there exists an ambiguity in the conductivity computing,
attributed to the wick behavior at ultraviolet vicinity. We argue
that to study electrical response of graphene completely,
non-perturbational contribution should be considered. \vskip 0.4in
\end{abstract}


\begin{multicols}{2}
\section{Introduction}
Graphene, a flat monolayer of carbon atoms tightly packed into a
two-dimensional honeycomb lattice, has spawned many theoretical and
experimental focuses. As stated in ref. \cite{0703742}, graphene
plays the role of bridge between condensed physics and high energy
physics. This is attributed to massless Dirac fermion behavior of
quasielectron in graphene, {\it i.e.} we can treat the
quasielectrons in graphene as ultimately relativity particles.

Such behavior also arouses many unusual properties of graphene,
such as ac and dc conductivity. Many
attentions\ct{mod,sigma1,sigma2,wallace} are on such topic.
However, there exist discrepancies in the problem, including the
discrepancy between theories and experiments on dc conductivity,
the famous missing "$\frac{1}{\pi}$ factor", and confliction among
different theoretical calculations\ct{ziegler}. So far different
theories are almost based on the perturbational approximation,
even calculations may be did by multi-loop diagrams\ct{multiloop}.

We introduce a correlation function with respect to only one
variable, the invariant amplitude of spatial-time position, $x$, to
study the graphene conductivity non-perturbationally. From the
correlation function, we show there is relation between imaginary
part and real part of ac conductivity. The function is very close to
spectral function and we find that the perturbational calculations
to conductivity only include contributions from free
valence-conduction electron pairs. Therefore, besides these
contributions, to compute conductivity completely we should also
consider other ones, such as excitation or impurity. To check the
statement, we perform a perturbational calculation of dc
conductivity using the quantum field theory. This technique
guarantees that the formulation is Lorentz-covariant and
local-gauge-invariant. We point out that there exists discrepancy
among different theoretical calculations, attributed to bad
behaviors at ultraviolet vicinity of $\delta$-functions.

We organize the paper as following: In section 2, we discuss the
electrical response to arbitrary external electric field. A
discussion on obtaining dc conductivity utilizing Kubo
theory\ct{kubo} is also given. The relation between imaginary part
and real part of ac conductivity is listed in section 3. We show an
explicit perturbational computing for conductivity in section 4 and
a brief discussion in section 5.

\section{The conductivity under arbitrary external fields}
To perform the calculation we first give the second quantization
on graphene briefly. Lagrangian density is  \bee
\label{lagrangian} \mathcal{L}=\bar{\psi} (i\gamma_\mu\partial^\mu
-m)\psi, \ee where $\partial^\mu=\frac{\partial}{\partial x_\mu}$,
$\bar{\psi}=\psi^\dag\gamma_0,$  $m$ is the mass of quasiparticle
(To clarify we here endow the quasielectrons with a nonzero mass),
and $\gamma$'s are  \bee \nonumber \gamma_0 = \beta= \tau_3,\,
\gamma_1 =\beta \tau_1,\, \gamma_2 =\beta\tau_2, \ee where
$\tau_1, \tau_2$ and $\tau_3$ are three Pauli matrices. In the
paper, the repeated indices are generally summed, unless otherwise
indicated. Furthermore, $\hbar=v_F=e=1$ are always set.

The Hamiltonian is then\ct{wallace} \bee H=\int d^2\mathbf{r}
\psi^\dag(\mathbf{r})\beta (-i
\gamma_i\partial^i+m)\psi(\mathbf{r}). \label{hamiltonian}\ee
Here, Latin indices $i,\,j,$ etc. generally run over two spatial
coordinate labels,1,2. (While Greek indices $\mu,~\nu,$ and so on,
three spatial-time coordinate labels 0, 1, 2 with $x_0$ the time
coordinate.)

Under the second quantization, we have \beea \psi(\vx)=\int
\frac{d^2\vp}{(2\pi)^2\sqrt{2p_0}}[a_\vp u(\vp)e^{-ipx}+b^\dag_\vp
v(\vp)e^{ipx}], &&\nonumber \\
\bar{\psi}(\vx)=\int \frac{d^2\vp}{(2\pi)^2\sqrt{2p_0}}[a^\dag_\vp
\bar{u}(\vp)e^{ipx}+b_\vp \bar{v}(\vp)e^{-ipx}], \nonumber \eea
where $p_0=\sqrt{m^2+\vp^2}>0$. Solutions to positive energy
$u(\vp)$ and to negative energy $v(\vp)$ satisfy respectively
\beea && u^\dag u=2p_0, \bar{u}u=2m, u\bar{u}=\sd{p}+m, \nonumber \\
&& v^\dag v=2p_0, \bar{v}v=-2m, v\bar{v}=\sd{p}-m
. \label{relation} \eea The explicit forms of $u$ and $v$ are
irrelevant. For simplification, we call operator
$a^\dag_\vp$($a_\vp$), which creates(annihilates) quasielectron in
conduction band, as creation(annihilation) operator which
creates(annihilates) electron, at the same time, we call operator
$b^\dag_\vp$($b_\vp$), which annihilates(creates) electron in
valance band, as creation(annihilation) operator which
creates(annihilates) hole. Both electron and hole have positive
energy, $p_0$.

Since in the perturbational ground state, the valance band is
completely filled while the conduction band is empty, the energy of
the ground state is nonzero, $E_{gnd}=-\int
d^2\vp\sqrt{m^2+\mathbf{p}^2}$. To obtain a Lorentz invariant ground
state, we perform a substraction for all the states, $E\rightarrow
E-E_{gnd}$. Under such subtraction, each physics quantity, such as
energy, current, etc. should be in normal form\ct{weinberg}.

With the substitution $i\partial_\mu\rightarrow p_\mu$, we read
Hamiltonian operator eventually \bee H=\int d^2\vp p_0(a^\dag_\vp
a_\vp+b^\dag_\vp b_\vp). \ee

When concerning of electromagnetic interactions we should make a
substitution of $p_\mu \rightarrow p_\mu-eA_\mu$ in Eq.
(\ref{hamiltonian}). Denoting $A_\mu=g_{\mu\nu}A^\nu$ with metric
matrix $g=diag\{1,-1,-1\}$, the interacting Lagrangian density is
$\mathcal{L}_{int}=-e\bar{\psi}\gamma_\mu\psi A^\mu=-J_\mu A^\mu$
and the corresponding Hamiltonian is \bee H_{int}=-\int d^2\vx
\mathcal{L}_{int}=\int d^2\vx J_\mu(\vx) A^\mu(\vx), \label{h-int}
\ee where, just as pointed out above, $J_\mu(x)$ is in norm form:
\beea && J_\mu(x)=:\bar{\psi}\gamma_\mu\psi(\vx): \nonumber \\
&& =\int \frac{d^2\vp
d^2\vp^\prime}{(2\pi)^4\sqrt{2p_0p_0^\prime}} \{ a^\dag_\vp
a_\vp^\prime e^{i(p-p^\prime)x} \bar{u}(\vp)\gamma_\mu
u(\vp^\prime) \nonumber \\ && +a^\dag_\vp b^\dag_{\vp^\prime}
e^{i(p+p^\prime)x} \bar{u}(\vp)\gamma_\mu v(\vp^\prime) \nonumber
\\ && +b_\vp a_{\vp^\prime} e^{-i(p+p^\prime)x}
\bar{v}(\vp)\gamma_\mu u(\vp^\prime) \nonumber \\
&& -b^\dag_{\vp^\prime} b_\vp e^{-i(p-p^\prime)x}
\bar{v}(\vp)\gamma_\mu v(\vp^\prime) \}. \nonumber \eea Unlike
some papers, we here introduce a factor ${1\over \sqrt{2p_0}}$
associated with momentum integration, which is attributed to the
Lorentz covariant\ct{weinberg,peskin}.

Both the electron number and the hole number are conservative
without interaction. However, the only conservation quantity is
their difference when interactions are included, \bee N=\int
d^2\vx :\psi^\dag\psi:=\int \frac{d^2\vp}{(2\pi)^2} (a^\dag_\vp
a_\vp-b^\dag_\vp b_\vp). \ee

Generally, the interacting Hamiltonian of graphene in external
field $A^\mu(x_0,\vx)$ is described by Eq. (\ref{h-int}). The
density operator is $\rho=\rho_0+\delta\rho$, where $\rho_0$ is
the equilibrium density operator and $\delta\rho$ is the leading
order correction with respect to the external field.

In Heisenberg picture we have \bee
i\delta\rho(x_0)=[H_{int},\rho_0].\ee Therefore\ct{mod}, \bee
\delta\rho(x_0)=-i\int^{x_0}_{-\infty} dx_0^\prime d\vx^\prime
[J_\mu(x_0^\prime\vx^\prime),\rho_0]A^\mu(x_0^\prime\vx^\prime).\ee

Due to the spatial and time translation invariant, at zero
temperature, on fixed time $x_0^0$, the current density at arbitrary
position is \beea &&
<J_\mu(x_0^0)>=Tr(\delta\rho(x^0_0)J_\mu(x^0_0\mathbf{0})) \nonumber
\\ && =\int^0_{-\infty}dx_0 d^2\vx
T_{\mu\nu}(x_0\vx)A^\nu(x_0+x_0^0\vx), \label{J-mu} \eea where \bee
T_{\mu\nu}(x_0\vx)=i<|[J_\nu(x_0\vx),J_\mu(0\bm{0})]|> \label{T-mu}
\ee with the notation of the ground state $|>$.

Noticing in Eq. (\ref{J-mu}) and (\ref{T-mu}), variable $x_0$ is
defined on $(-\infty, 0)$, we expand the range of $x_0$ onto
$(-\infty, \infty)$ as \beea
&&T_{\mu\nu}(x)=i\theta(x_0)<|[J_\mu(0),J_\nu(x)]|> \nonumber
\\ && ~~~~+i\theta(-x_0)<|[J_\nu(x),J_\mu(0)]|>, \label{T-define}\eea
where $\theta(x_0)$ is step function: $\theta(x_0)=1$ for $x_0
\geq 0$,$\theta(x_0)=0$ for $x_0 \leq 0$. Tensor $T_{\mu\nu}(x)$
is vanishing for space-like $x$, {\it i.e.} the support of tensor
$T_{\mu\nu}$ is time-like three-dimensional vector $x$.

Our expansion is different to the one in Ref. \cite{mod}, where
$T_{\mu\nu}$ has only forward term or backward term.

The conductivity should be independent on the gauge transformation.
This means that, under a local gauge transformation
$A^\nu\rightarrow A^{\prime\nu}=A^\nu-\partial^\nu f$, where $f$ is
arbitrary function with $f(x_0=x_0^0)=f(x_0=-\infty)=0$, the current
density in Eq. (\ref{J-mu}) should be invariant. Integrating Eq.
(\ref{J-mu}) by part we find that this requirement is satisfied
provided $\partial^\nu T_{\mu\nu}=0$ everywhere. The statement can
be proven by the facts: 1)charge conversation, {\it i.e.}
$\partial^\nu J_\nu\equiv 0$; 2)${\partial \over
\partial x_\mu}\theta(x_0)=-{\partial \over
\partial x_\mu}\theta(-x_0)=\delta_{0\mu}\delta(x_0)$; 3) the equal time
commutation relation $[J_0(t\vx),J_\mu(t\vy)]=0$.

$T_{\mu\nu}(x)$ is written as \bee
T_{\mu\nu}(x)=-(\partial_\mu\partial_\nu-g_{\mu\nu}\square) \Pi(x),
\ee where $\square\equiv
\partial^\mu\partial_\mu=\frac{\partial^2}{\partial x_0^2} -
\frac{\partial^2}{\partial x_1^2}- \frac{\partial^2}{\partial
x_2^2}$, $\Pi(x)$ is scalar function with respect to only one
variable, invariant amplitude of three-dimensional spatial-time
vector $x$. After defining the Fourier transformation of function
$f(x)$ as $ f(q)=\int d\vx dt f(x)e^{i\, q\, x}$ with $q\,x\equiv
q_\mu x^\mu$, we have, then, \bee T_{\mu\nu}(q)=(q_\mu
q_\nu-q^2g_{\mu\nu})\Pi(q^2), \ee where $\Pi(q^2)$ is the only
function with respect to invariant amplitude of $q$.

From Eq. (\ref{J-mu}) it seems that $J_{\mu}$ is time dependent in
the time-invariant external electric field. But this is not true.
It is enough to  illustrate it by a special gauge,
$A^\nu=(0,E\,x^0,0)$ or $A_\nu=(0,-E\,x_0,0)$, where $A$ is
three-dimensional potential and $E$ is the external electric
field, $E_\nu =(E, 0)$. It is easily to see that
\bee  J_\mu(x_0^0)
=-E\int_{-\infty}^0 \frac{dx_0 dq_0}{2\pi}e^{-i\,q_0\,x_0}g_{\mu
1}q^2_0\Pi(q_0)(x_0+x_0^0),   ~~~ \ee where $\Pi(q_0)\equiv
\Pi(q_0^2,\mathbf{q}=\mathbf{0})$. $J_0$(charge density) and $J_2$
are both vanishing and only $J_1\neq 0$: \beea J_1(x_0^0)&=&E
\int{dq_0\over 2\pi}q_0^2\Pi(q_0^2) \int^{x_0^0}_{-\infty}dx_0 x_0
e^{iq_0x_0+\epsilon x_0} e^{iq_0x_0^0} \nonumber \\
&=&E\int {dq_0\over 2\pi}
\Pi(q_0^2)\frac{q_0^2}{q_0^2+iq_0\epsilon}(1+iq_0x_0^0)\nonumber \\
&\equiv&E\sigma(x_0^0), \label{j-sigma}\eea where the additional
factor $e^{\epsilon x_0}$ ($\epsilon$ is a positive infinitesimal)
is to guarantee that the external field is introduced adiabatically.
When $\epsilon\rightarrow 0$, $\frac{q_0^2}{q_0^2+iq_0\epsilon}
\rightarrow 1$, we can replace $\frac{q_0^2}{q_0^2+iq_0\epsilon}$ by
unitary. Furthermore, since $\Pi(q_0^2)$ is even function of $q_0$,
$\int {dq_0\over 2\pi} \Pi(q_0^2) iq_0\,x_0\equiv 0$. We finally
have a time-independent conductivity \bee \sigma=\int {dq_0\over
2\pi}\Pi(q_0^2)\equiv \Pi(x_0=0,\mathbf{q}=\mathbf{0}),
\label{sigma} \ee where, to obtain a meaningful quantity, we should
perform a substraction of $\Pi$, {\it i.e.} we make a substitution:
$\Pi(x_0=0,\mathbf{q}=\mathbf{0}) \rightarrow
\Pi(x_0=0,\mathbf{q}=\mathbf{0})-\Pi(x_0=-\infty,\mathbf{q}=\mathbf{0})$.
In Fourier space, this substraction is the substitution
$\Pi(q_0,\mathbf{q}) \rightarrow
\Pi(q_0,\mathbf{q})-\Pi(q_0=0,\mathbf{q})$. In the paper, we always
make such substraction for all the physical quantities.

As expected, we obtain a time-independent current density for a
steady external field.

It is not difficult to deduce the response to arbitrary external
fields. Supposing the external electric field is ac with frequency
$\omega$, $A=(0,E_0\,e^{i\omega
 x_0},0)e^{\epsilon x_0}$ and substituting the potential into
expression (\ref{J-mu}), we find that only x-component of current
density is nonvanishing,
 \bee J_1(x_0)=E_0 e^{i\omega x_0}\int{dq_0\over
 2\pi}\Pi(q_0^2,\mathbf{0})\frac{i q_0^2}{q_0-\omega+i\epsilon}. \ee
Since this potential stands for external electric field
$(E_1,E_2)=(i\omega\,E_0e^{i\omega x_0},0)$, the complex
conductivity is \begin{subequations}\beea \sigma=&\int{dq_0\over
 2\pi}\Pi(q_0^2,\mathbf{0})\frac{q_0^2}{\omega(q_0-\omega+i\epsilon)}
 ~~~~~~~~~~~~~~~~~~~~~\label{sigma2} \\
=&{1\over \omega}\int{dq_0\over
 2\pi}\Pi(q_0^2,\mathbf{0})q_0
+\int{dq_0\over
 2\pi} \Pi(q_0^2,\mathbf{0}){q_0\over q_0-\omega+i\epsilon} .
 \label{sigma3}
 \eea \label{sigma23} \end{subequations}
Obtaining the dc conductivity from Eq. (\ref{sigma2}) and
(\ref{sigma3}) corresponds to the results obtained from famous
Kubo theory. However, it is not obvious whether we can obtain dc
conductivity (\ref{sigma}) from the limit of Eq. (\ref{sigma2}):
Firstly, to obtain Eq. (\ref{sigma3}) from (\ref{sigma2}) we need
not only convergence of all the integrations, such as $\int
{dq_0\over 2\pi}
\Pi(q_0^2,\mathbf{0})\frac{q_0^2}{\omega(q_0-\omega+i\epsilon)}$,
etc. but also proper subtraction of physics quantities. Secondly,
when one chooses $\omega=0$ directly in Eq. (\ref{sigma3}), he
will meet an uncomfortable situation: the first term in Eq.
(\ref{sigma3}) is an ambiguous ${0\over 0}$. To obtain right dc
conductivity we should perform computation as follows: we
calculate the ac conductivity in the course of nature from Eq.
(\ref{sigma23}) at $\omega\neq 0$, with proper subtraction. At
last we read the dc conductivity utilizing the limit of
$\omega\rightarrow 0$. This is just pointed out by Kubo\ct{kubo},
which implies that, compare to results in references
\cite{sigma1}, results in references \cite{sigma2} is just the
right results. Of course, all the results in references
\cite{sigma1,sigma2} are obtained by perturbational approach.

After suitable subtraction, the ac conductivity is \bee
\label{acsigma} \sigma=\int{dq_0\over
 2\pi} \Pi(q_0^2,\mathbf{0}){q_0\over q_0-\omega+i\epsilon} \,.
 \ee
Generally, $\Pi(q_0^2,\mathbf{0})$ is not convergent or well
defined, as will be shown by perturbative calculation in section 4.
This is relevant to the wick definition of $\delta-$function.
Therefore, in order to obtain meaningful physical result, we need to
perform subtraction to cancel divergence.  For instance, in Ref.
\cite{ziegler} the author has proposed a soft $\delta-$function. The
subtraction should meet some physical criteria. For instance, as
found in above paragraph, to get the result in Eq. (\ref{acsigma})
from Eq. (\ref{sigma23}), after the subtraction $\Pi$ is still the
function of $q_0^2$ rather than the function of $q_0$ in Fourier
space. In section 4 we shall show a explicit subtraction to
$\Pi(x_0,\vq)$.

\section{Kramers-Kr\(\ddot{\mbox{o}}\)nig relation of graphene conductivity}
In this section we show a relation between imaginary part and real
part of graphene conductivity.

We first give a non-perturbational proof that $\Pi(q^2)$ is real.
$\Pi(q^2)$ is real at $q^2<0$ obviously, the only needed is to prove
that $\Pi(q^2)$ is also real at $q^2>0$.

At $q^2>0$, after inserting complete intermediate states
$\sum\limits_\Gamma|\Gamma><\Gamma|$, we have for $T\equiv
T^\mu_\mu(q)$, \bee  T=\int
\frac{d^3xd^3pe^{iqx}}{(2\pi)^2}i(\theta(x_0)-\theta(-x_0)) (e^{ip
x}-e^{-ip x})s(p)\theta(p_0),~~~ \nonumber\ee where the spectral
function $s(p)$ is defined as \bee 2\pi s(p)=\sum\limits_\Gamma
<|J^\mu(0)|\Gamma><\Gamma|J_\mu(0)|>(2\pi)^3\delta^3(p-p_\Gamma).
\ee

The spectral function $s(p)$, which is very close to state density,
includes not only perturbational contributions, but also
non-perturbational contributions. To study non-perturbational
contributions, one should consider, for instance, excitations.

Inserting $\int_0^\infty dt\delta(p^2-t)\equiv 1$ at $p^2>0$, one
obtains \bee T=\int_0^\infty dt s(t)\int
d^3xe^{iqx}i(\theta(x_0)-\theta(-x_0))I, \ee where \beea && I=
\int{d^3p\over (2\pi)^2}\theta(p_0)\delta(p^2-t)(e^{ip x}-e^{-ip
x}) \nonumber \\ && =\int{d^3p\over
(2\pi)^2}e^{-ipx}\delta(p^2-t)(\theta(-p_0)-\theta(p_0)). \notag
\eea Utilizing \beea && i\int {d^3pe^{-ipx}\over
(2\pi)^2}[\theta(x_0)\theta(p_0)+\theta(-x_0)\theta(-p_0)]\delta(p^2-t)
\notag \\ && =-\int{d^3p\over
(2\pi)^3}\frac{e^{-ipx}}{p^2-t+i\epsilon}, \notag \eea \beea &&
i\int {d^3pe^{-ipx}\over
(2\pi)^2}[\theta(-x_0)\theta(p_0)+\theta(x_0)\theta(-p_0)]\delta(p^2-t)
\notag \\ && =\int{d^3p\over
(2\pi)^3}\frac{e^{-ipx}}{p^2-t-i\epsilon},\notag \eea we finally
get \bee T=2P\int_0^\infty dt s(t){1\over q^2-t} \label{tq} \ee
with identity ${1\over f\pm i\epsilon}=P{1\over f}\mp
i\pi\delta(f)$. Since $s(t)\geq 0$ is real, $T(q)$ and therefore
$\Pi(q)$ are both real. The spectral density is in fact a very
important function, which will be studied elsewhere\ct{lz}.

Thus, from Eq. (\ref{acsigma}), we obtain an important relation
between real part and imaginary part of conductivity $\sigma$ \bee
\mbox{Im}\sigma(\omega)=-{\omega\Pi(\omega^2,\mathbf{0})\over 2},\,
\mbox{Re}\sigma(\omega) =P\int {ds\over
\pi}\frac{\mbox{Im}\sigma(s)}{\omega-s}. \label{im-real} \ee This
relationship between real part and imaginary part of conductivity is
beyond the perturbation approach and can be considered as
Kramers-Kr\(\ddot{\mbox{o}}\)nig relation of graphene conductivity.
We hope the advanced study of graphene may check this relation.

Eq. (\ref{im-real}) is one of the main results of the paper. It
points out that the electrical response of graphene can never be
considered as a pure resistance, but a resistance parallel connected
with a capacitor with capacitivity $\Pi(\omega^2,\mathbf{0})/2$.
Furthermore, due to the obvious relation between $\mbox{Im}\sigma$
and $\Pi$, $\mbox{Im}\sigma$ reflects the state structure of
graphene. $\mbox{Im}\sigma$ is therefore a non-perturbational probe
to detect the state structure of graphene. In this viewpoint,
$\mbox{Im}\sigma$ is a more basic quantity than $\mbox{Re}\sigma$.
Furthermore, Eq. (\ref{im-real}) is irrelevant to the idiographic
interactions, which means that, the equation holds under very
general conditions, such as the existence of impurities or
excitations in graphene.

More present works reveal that the graphene is rarely flat, {\it
i.e.} there are always ripples in graphene. The nonvanishing
curvature, raised by ripples, will lead two main effects: altering
group velocity of quasiparticle and introducing effective gauge
fields. The first effect possibly makes a global correction to
conductivity, which may be absorbed into the redefinition of
spectral function, $\Pi$. Furthermore, since the holding of Eq.
(\ref{im-real}) is irrelevant to idiographic interactions, we
conclude that Kramers-Kr\(\ddot{\mbox{o}}\)nig relation is still
valid for corrugated graphene.

Eq. (\ref{im-real}) supplies one possible way to study the
discrepancy of dc conductivity between theories and experiments. One
may first perform perturbational computing to imaginary part of ac
conductivity and then compares the perturbational result to
experiments at different frequency. The discrepancy between these
results reveals contribution which can not be ascribed to
perturbational theory. Furthermore, the complete contributions to dc
conductivity are not only from the Dirac nodal point, but from the
spectral structure of carriers.

\section{Perturbational Calculation of dc conductivity}
We here deduce the dc conductivity in perturbational approximation.
After that we shall discuss an ambiguity besides the one pointed out
in Ref. \cite{ziegler}.

We begin the game by a perturbational calculation of $T_{\mu\nu}$.
Noticing $<0|b_{\vp_f}a_{\vp_f^\prime}a_{\vp_i}^\dag
b_{\vp_i^\prime}^\dag |0>=(2\pi)^4
\delta^2(\vp_f-\vp_i^\prime)\delta^2(\vp_f^\prime-\vp_i)$ and the
normal ordering of operators in current density, we have, for
perturbational ground state, \beea <0|J^\mu(x)J_\mu(0)|0>=\int
\frac{d^2\vp d^2\vp^\prime}{(2\pi)^4 2p_02p_0^\prime}F_\mu^\mu
e^{-i(p+p^\prime)x}, \eea  where
$F_\mu^\mu=\bar{v}(\vp^\prime)\gamma^\mu u(\vp)
\bar{u}(\vp)\gamma_\mu v(\vp^\prime)$. Taking advantage of Eq.
(\ref{relation}), one finds, \bee F^\mu_\mu=-2p\cdot
p^\prime-6m^2=-(p+p^\prime)^2-4m^2. \ee $T^\mu_\mu\equiv T$ is given
by a direct computing \bee
T(x)=i(\theta(x_0)-\theta(-x_0))(\square-4m^2)K^\prime(x), \ee where
\beea K^\prime(x)=\int \frac{d^2\vp d^2\vp^\prime}{(2\pi)^4
2p_02p_0^\prime}(e^{i(p+p^\prime)x}-e^{-i(p+p^\prime)x}) ~~~~~~
&&\nonumber
\\
=\int \frac{d\vp (e^{ipx}+e^{-ipx})}{(2\pi)^2 2p_0} \int \frac{d\vp
(e^{ipx}-e^{-ipx})}{(2\pi)^2 2p_0} && \nonumber
 \eea
From
\beea \int \frac{d\vp(e^{ipx}+e^{-ipx})}{(2\pi)^2
2p_0}=\int {d^3pe^{-ipx}\over (2\pi)^2}\delta(p^2-m^2), ~~~~ \notag \\
\int \frac{d\vp(e^{ipx}-e^{-ipx})}{(2\pi)^2 2p_0}=\int
{d^3pe^{-ipx}\over (2\pi)^2}\delta(p^2-m^2)\mbox{sgn}(p_0),
\nonumber  \eea Fourier transformation of $K^\prime$ is \beea
K^\prime(p)=-\int {d^3q \over 2\pi}
\delta((p-q)^2-m^2)\delta(q^2-m^2)\mbox{sgn}(q_0). \eea Here $p$
does not need to be on mass shell, {\it i.e.}
$p_0=\sqrt{\vp^2+m^2}$ is not needed, if the integrating factor is
$d^3p$. We focus on the case of $\vp=0$(Or, $p$ is a time-like
vector). Letting $K(x)=(\square-4m^2)K^\prime(x)$ and
$K(q_0)=K(q_0,\mathbf{q}=\mathbf{0})$, we get \bee
K(q_0)=\frac{q_0^2+4m^2}{4q_0}\theta(q_0^2-4m^2). \label{kq0} \ee
The nonzero contribution to $K(q_0)$ is $|q_0|>2m$. To simplify we
let $m=0$. Thus $T(x_0,\vp=0)={1\over
4}(\theta(-x_0)-\theta(x_0))\delta^\prime(x_0)$. Since
$\Pi(q_0,\vq=0)=-{1\over 2q_0^2}T(q_0,\vq=0)$, we find the dc
conductivity of graphene \bee \sigma={1\over
8}\int^0_{-\infty}dx_1
\int^{x_1}_{-\infty}dx_2(\theta(-x_2)-\theta(x_2))\delta^\prime(x_2),
\label{pi2} \ee utilizing Eq. (\ref{sigma}). Notice that in above
equation we have made a subtraction
$\Pi(x_0,\vp)=\frac{\partial\Pi(x_0,\vp)}{\partial x_0}=0$ at
$x_0\rightarrow -\infty$.

However, the functions, such as $\delta^\prime(x)$ and
$\theta(x)$, are not well defined. This means that there possibly
exists ambiguity in Eq. (\ref{pi2}). This ambiguity is different
to the one pointed out in Ref. \cite{ziegler}.

We consider dc conductivity here. First let $\delta(x)$ be the
simplest form, $\delta_1(x)=0$ for $|x|>{a\over 2}$ and
$\delta_1(x)=1/a$ for $|x|<{a\over 2}$. In this case we obtain
$\sigma_1={1\over 16}={\pi\over 8}{1\over 2\pi}\simeq {0.39\over
2\pi}$ utilizing Eq. (\ref{pi2}). This is just the result obtained
in Ref. \cite{sigma2}. Meanwhile, we can also let $\delta(x)$ be a
somewhat complex form\ct{ziegler}, $\delta_2(x)={1\over
\pi}{\eta\over x^2+\eta^2}$. At this time we get
$\sigma_2={4+\pi^2\over 16\pi}{1\over 2\pi}\simeq {0.28\over
2\pi}$. Finally, we can also set $\delta_3(x)={1\over
4T_0}\cosh^2({x\over 2T_0})$. We find $\sigma_3={\pi(1/2+\ln
2)\over 12}{1\over 2\pi}\simeq {0.31\over 2\pi}\simeq {1\over
\pi}{1\over 2\pi}$, numeral value of which is in agreement with
that in ref. \cite{sigma1}.

To see the physics meaning of $T_0$ in $\delta_3(x)$, we write out
explicitly: $\theta(-x)={1\over 1+e^{t/T_0}}$. The role of $T_0$
is some like temperature, which means that $T_0^{-1}$ symbolizes
the disorder. $a^{-1}$ in $\delta_1(x)$ and $\eta$ in
$\delta_2(x)$\ct{ziegler} play the similar role. Since
$\sigma_1,\,\sigma_2$ and $\sigma_3$ are $a-,\eta-$ and
$T_0$-independent respectively, we conclude that the dc
conductivity is almost temperature-independent near zero
temperature, although the conductivity value is ambiguous because
of the wicked behavior of $\delta$-function. This is verified by
experments\ct{nature}.

This is a unexpected occasion that the conductivity, a physical
observable quantity, varies with different definitions of
$\delta-$function. The ambiguity is associated by the different
definitions of $\delta-$function at ultraviolet region. One may
argue that we can eliminate the ambiguity by a standard
renormalization schedule in quantum field
theory\ct{weinberg,peskin}, however, this elimination is still
contributed to the special definition of $\delta-$function at
ultraviolet region. We think that the ambiguity implies that the
dc conductivity of graphene depends on the behavior of
quasielectrons at high energy as well as the behavior at Dirac
nodal point. This is also pointed by
Kramers-Kr\(\ddot{\mbox{o}}\)nig relation in Eq. (\ref{im-real}).
Unfortunately, linear dispersion relation of quasielectron does
not hold at high energy, which means that, different numeral
values based on linear dispersion and perturbational approaches,
need corrections. On the other hand, when we study electrical
response of graphene, we always perform calculations utilizing
diagrams composed by different Green functions. To include higher
corrections, we should use loop diagrams. However, since coupling
$g=2\pi e^2/\epsilon \hbar v_F$ is not small, $g\sim 1$, comparing
to leading order, the loop corrections can not be ignored.

One possibly expects that the correction to conductivity given above
are not large. If this is the case, our computations and
others\ct{sigma1,sigma2} indicate that about 30\% of the full
conductivity is from the perturbational contribution. A question is
raised, then, where other contributions to conductivity come from. A
generalized version of Eq. (\ref{kq0}) tells us that, from the
definition of state density $s(q)$, perturbational contribution to
state density is \bee 2\pi
s^{pt}(q)=\frac{q^2+4m^2}{4q}\theta(q^2-4m^2), \label{state-density}
\ee at $q_0>0$. $\theta$ function in this equation reveals that,
$s^{pt}$ only includes the contribution from pairs of {\it free}
quasielectron and hole. However, since there are complex
interactions between electron and hole, electron and hole may be
combined into excitations\ct{exciton,plasmon}, or in other words, it
is questionable to consider quasielectrons in graphene as
2-dimensional electron gas with no interacting. To study electrical
responses completely, one must also consider the contribution of
excitations(and impurities), attributed to Eq. (\ref{tq}). In
standard field theory it is difficult to study the contribution
perturbationally. We often nominate the contribution as
non-perturbational one, such as we did in Ref. \cite{shifman}. Since
the coupling is large on graphene, such contribution can not be
ignored when one consider electrical responses. Apparently, if $m$
is large enough, the nonzero contribution from exciton appears
before $q^2=4m^2$. We shall discuss such contribution
elsewhere\ct{lz}.

\section{Discussion}
The relationship between imaginary part, $\mbox{Im}\sigma$, and
real part, $\mbox{Re}\sigma$, of ac conductivity is given in
paper. $\mbox{Im}\sigma$ depends directly on details of state
structure and one can study state structure from
$\mbox{Im}\sigma$. We consider it as a non-perturbational probe to
detect state structure of graphene and it is therefore a very
important quantity. Our formulae are Lorenz-covariant and
local-gauge-invariant.

We also perform an explicit perturbational calculation using quantum
field theory. The computing shows that the conductivity is mainly
manipulated by the momentum-energy relation and there is little
nexus between the conductivity and state density near Dirac nodal
point. The computing reveals that, due to the wicked behavior of
$\delta$-function, there is ambiguity in graphene conductivity
calculations. We argue that the full perturbational studies need two
corrections: one is due to the incorrectness of carrier linear
dispersion at high-energy and the other is higher order correction.
Besides these corrections, however, there is a furthermore
correction which is nominated as non-perturbational corrections in
the paper. This correction comes from the contribution of
excitations, which is attributed to electron-electron interactions.

Authors are very grateful to Dr. M.G. Xia and Dr. E.H. Zhang. This
work is supported by the Ministry of Science and Technology of China
through 973 - project under grant No. 2002CB613307, the National
Natural Science Foundation of China under grant No. 50472052 and No.
60528008.

\end{multicols}
\end{document}